\documentclass[final,english,twocolumn,amssymb, 
               nobibnotes,aps,prl,longbibliography]{revtex4-1}
\usepackage{graphicx}
\usepackage{natbib}
\newcommand{\tr}{\mathrm{Tr}}
\usepackage{amsmath}
\usepackage{amssymb}
\usepackage{bbold}
\usepackage[caption=false]{subfig}
\usepackage{float}
\usepackage{appendix}
\usepackage[dvipsnames]{xcolor}
\usepackage[section]{placeins}
\definecolor{darkblue}{rgb}{0,0,.65}
\definecolor{darkgreen}{rgb}{0.28,0.41,0.19}
\usepackage{wrapfig}

\usepackage[%
pdfstartview=FitH,%
breaklinks=true,%
bookmarks=true,%
colorlinks=true,%
anchorcolor=black,%
citecolor=blue,
filecolor=black,%
menucolor=black,%
urlcolor=darkblue,%
linkcolor=blue,%
]{hyperref}
\usepackage[all]{hypcap} %
\newcommand{\bra}[1]{\langle\,#1\,|}
\newcommand{\ket}[1]{|#1\rangle}

\usepackage[normalem]{ulem}
\graphicspath{{figs/}}
\begin{document}

	\title{Compressed quantum error mitigation}
	
	\author{Maurits S. J. Tepaske}
	\affiliation{Physikalisches Institut, Universit\"at Bonn, Nu{\ss}allee 12, 53115 Bonn, Germany}
	\author{David J. Luitz}
	\email{david.luitz@uni-bonn.de}
	\affiliation{Physikalisches Institut, Universit\"at Bonn, Nu{\ss}allee 12, 53115 Bonn, Germany}

	\date{\today}
	
	\begin{abstract}
    We introduce a quantum error mitigation technique based on probabilistic error cancellation to eliminate errors
    which have accumulated during the application of a quantum circuit. 
    Our approach is based on applying an optimal ``denoiser'' after the action of a noisy circuit
    and can be performed with an arbitrary number of extra gates. 
    The denoiser is given by an ensemble of circuits distributed with a quasiprobability distribution.
    For a simple noise model, we show that efficient, local denoisers can be found, and we demonstrate 
    their effectiveness for the digital quantum simulation of the time evolution of simple spin chains.
	\end{abstract}
	\maketitle

\textit{Introduction.} --- 
Quantum information processing has been theoretically shown to hold great 
promises, and quantum algorithms were developed which can in principle achieve an exponential speed-up 
over their classical counterparts, both for general purpose computing 
\cite{kitaev_quantum_1995,shor_polynomial_1997,ebadi_quantum_2022, arute_quantum_2019} 
and quantum simulation \cite{feynman_simulating_1982,lloyd_universal_1996, barison2021efficient, banuls2020simulating, scholl2021quantum}. 
However, present day quantum computing prototypes still suffer from significant noise processes 
which hinder the execution of many potentially groundbreaking quantum algorithms \cite{preskill_quantum_2018}.
Nontrivial quantum algorithms typically require large sequences of quantum gates, each of 
which introduces dissipation and hence an overall loss of coherence, eventually rendering 
the results useless.

Until quantum error correction \cite{calderbank_good_1996, shor_scheme_1995} becomes practical, 
\textit{quantum error mitigation} seems to be more feasible to increase the accuracy of expectation values. 
Here the goal is to induce the (partial) cancellation of errors that stem from noisy quantum gates 
by extending the circuit corresponding to the desired algorithm 
with an ensemble of gates \cite{temme_error_2017, endo_practical_2018}, sampled from a quasiprobability distribution.

The traditional way to accomplish this is with the gate-wise method from \cite{temme_error_2017, endo_practical_2018}, 
where noise is mitigated by inverting the noise channel of each gate separately, i.e. the cancellation 
of errors is performed for each gate on its own. Here the local noise channel is 
approximated in a way such that it can be easily inverted analytically, e.g. using Pauli twirling \cite{endo_practical_2018}.
Gates are then sampled from the inverted noise channel by interpreting it as a quasiprobability distribution. 
Because in this gate-wise approach every noisy gate has to be modified separately, the sign problem 
is exponentially large in the number of gates, limiting the practicality of the mitigation. 
The success of the gate-wise approach resulted in a large body of 
work concerning these methods \cite{vovrosh_simple_2021, filippov_matrix_2022, cao_nisq_2021, piveteau_quasiprobability_2022, 
gutierrez_error_2016, jiang_physical_2021, magesan_modeling_2013,
cai_quantum_2022, ferracin_efficiently_2022}, including extensions for simultaneous mitigation of multiple gates 
by Pauli-twirling entire layers \cite{berg_probabilistic_2022, mcdonough_automated_2022} or variationally constructing 
a mitigating matrix product operator \cite{guo_quantum_2022}. 

In principle, errors during the execution of a circuit can propagate and accumulate. 
These propagated errors can potentially blow up and lead to large errors for the circuit as a whole \cite{flannigan_propagation_2022, poggi_quantifying_2020}.
Here we introduce a mitigation technique that takes into account the propagation of errors,  
can be performed with a tunable number of extra gates, and works for non-Clifford local noise channels 
since the inversion of the accumulated global noise channel is implicit. 
We first execute the targeted noisy circuit completely, letting the noise propagate and accumulate, 
and only afterwards we apply an extra random circuit sampled from a quasiprobability distribution.
We call the corresponding ensemble of random circuits a \textit{denoiser}, 
and we construct it such that upon averaging the accumulated errors cancel. Essentially, the 
denoiser inverts a global noise channel. Since we will construct it as a local brickwall circuit,
following the classical pre-processing approach from \cite{tepaske_optimal_2022}, we call this 
\textit{compressed} quantum error mitigation.

\begin{figure}
        \centering
        \includegraphics[width=\columnwidth]{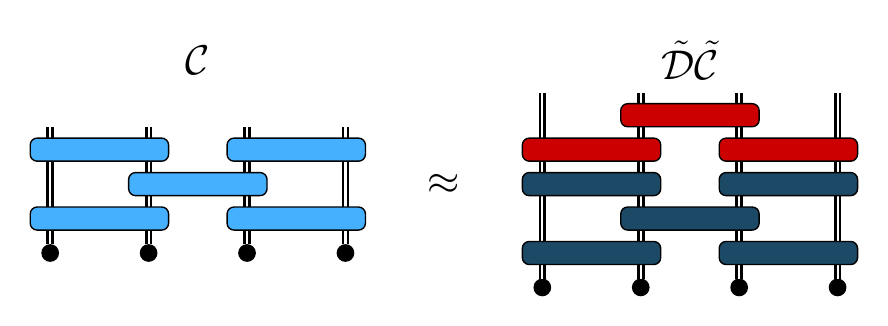}
        \caption{An example of the quantum error mitigation procedure used in this work for the time evolution of 
            the wave function of a spin chain. The ideal 
                 second-order Trotter supercircuit $\mathcal{C}$ of depth $M_{\text{trot}}=1$ (light blue) 
                 is approximated  by applying a denoiser $\tilde{\mathcal{D}}$ of depth $M=1$ (red) to the noisy Trotter supercircuit $\tilde{\mathcal{C}}$ (dark blue). 
                 Because the denoiser is applied 
                 after fully executing the noisy Trotter supercircuit, it represents an approximate inverse of the
                 global noise channel with a precision tunable by the depth of the denoiser.
                 }
        \label{denoise_circuit}
\end{figure}

\textit{Method.} --- 
Due to the inevitable coupling of a quantum processor to its environment, every qubit operation 
is affected by noise. Therefore, the simplest technique to minimize the impact of the resulting noise is to minimize the
number of operations when performing a quantum algorithm. In \cite{tepaske_optimal_2022} we showed that many-body time
evolution operators can be efficiently compressed into brickwall circuits with high fidelity per gate. 

In this Letter, we consider the noise explicitly by
treating quantum operations as (generally non-unitary) quantum channels, corresponding to completely positive and trace preserving (CPTP) maps \cite{nielsen_quantum_2011}. 
For example, instead of a noiseless two-qubit gate $G$, which acts on 
a quantum state $|\rho\rangle\rangle$ in superoperator form as $\mathcal{G}|\rho\rangle\rangle=G\otimes G^*|\rho\rangle\rangle$, we 
get the noisy channel $\tilde{\mathcal{G}}=\mathcal{N}\mathcal{G}$, where the noise channel 
$\mathcal{N}$ implements the two-qubit noise \cite{aharonov_quantum_1998}. 
These channels are used to construct a ``supercircuit'' $\tilde{\mathcal{C}}=\prod_{i=1}^{N_{\mathcal{G}}}\tilde{\mathcal{G}}_i$, 
consisting of $N_{\mathcal{G}}$ channels, which is affected by multi-qubit accumulated noise. This 
supercircuit encodes an ensemble of circuits \cite{aharonov_quantum_1998}. For simplicity, we 
assume that the noisy channels $\tilde{\mathcal{G}}_i$ in each half brickwall layer are lattice inversion and 
translation invariant, such that we can construct a denoiser with these properties, limiting the number of variational
parameters.

The purpose of quantum error mitigation is to modify the ensemble of circuits described by $\tilde{\mathcal{C}}$
in a way that we can use it to obtain the noiseless expectation values. 
In superoperator language, we do this by following the supercircuit $\tilde{\mathcal{C}}$ with a denoiser 
supercircuit $\tilde{\mathcal{D}}$, such that $\tilde{\mathcal{D}}\tilde{\mathcal{C}}$ is as close to the 
noiseless supercircuit $\mathcal{C}=C\otimes C^*$ as possible. Here $C$ is the target unitary circuit.
Because the noise channel $\mathcal{N}$ is non-unitary, hence making the supercircuit $\tilde{\mathcal{C}}$ non-unitary,
we need to use a non-unitary denoiser to retrieve the unitary $\mathcal{C}$.

We illustrate the mitigation procedure in Fig. \ref{denoise_circuit}, where a denoiser with one layer 
is used to mitigate errors for a second-order Trotter supercircuit with one layer. This circuit architecture 
is commonly used to simulate the time evolution of a quantum many-body system, until some time $t$, with controllable 
precision \cite{trotter_1959_product, suzuki_1976_generalized, paeckel_time_2019, tepaske_optimal_2022, ostmeyer_optimised_2022, 
kargi_2021_quantum, heyl_2019_quantum, childs_2021_theory, hemery_matrix_2019, 
variational_mansuroglu_2021, zhao_making_2022, berthusen_quantum_2021}, and we will use it to benchmark the denoiser. 
In practice, we cannot directly implement a supercircuit, and so we have to utilize its interpretation as an
ensemble of circuits. Essentially, after executing a shot of the noisy circuit we sample 
the denoiser and apply it. The goal is to construct the denoiser in a way that averaging 
over many of its samples cancels the accumulated errors and gives us a good approximation of the 
noiseless expectation values. 

It should be noted that our approach requires more gate applications on the 
quantum processor than with the gate-wise scheme, since there each sample from the mitigation 
quasiprobability distribution can be absorbed into the original circuit, whereas our approach 
increases the circuit depth.
We take this into account by imposing the same noise on the denoiser. Furthermore, within our scheme, the dimensionality 
of the quasiprobabilistic mitigating ensemble can be controlled, in contrast to the gate-wise approach 
where it is equal to the gate count.

To facilitate the stochastic interpretation we parameterize each 
two-qubit denoiser channel $\mathcal{G}_i$ as a sum of CPTP maps, such that we can sample the 
terms in this sum and execute the sampled gate on the quantum processor. Concretely, we use a trace preserving
sum of a unitary and a non-unitary channel. For the unitary part we take a two-qubit unitary channel 
$\mathcal{U}(\vec{\phi}_i)=U(\vec{\phi}_i)\otimes U^*(\vec{\phi}_i)$, with $U(\vec{\phi}_i)$ a two-qubit 
unitary gate parameterized by $\vec{\phi}_i$. For this we take the 
two-qubit ZZ rotation $\exp(-i\alpha(\sigma_z\otimes\sigma_z))$ with angle $\alpha$, which can be obtained from native
gates on current hardware \cite{chen_error_2022}, and dress it 
with four general one-qubit unitaries, only two of which are independent if we want a circuit that is 
space inversion symmetric around every bond. The resulting gate has 7 real parameters $\vec{\phi}_i$.

For the non-unitary part, which is essential because $\tilde{\mathcal{D}}$ has to cancel the non-unitary accumulated 
noise to obtain the noiseless unitary circuit, we use a general one-qubit measurement followed by conditional 
preparation channel $\mathcal{M}(\vec{\zeta}_i)|\rho\rangle\rangle=\sum_l K_l\otimes K_l^*|\rho\rangle\rangle$. It has Kraus operators 
$K_1=\ket{\psi_1}\bra{\psi}$ and $K_2=\ket{\psi_2}\bra{\bar{\psi}}$ if we measure in the orthonormal basis 
$\{\ket{\psi},\ket{\bar{\psi}}\}$, where $\ket{\bar{\psi}}$ is uniquely defined by $\ket{\psi}$ 
as they are antipodal points on the Bloch sphere.  
If the measurement yields $\ket{\psi}$ we prepare $\ket{\psi_1}$ and if we 
measure $\ket{\bar{\psi}}$ we prepare $\ket{\psi_2}$. These states can be arbitrary points on the 
Bloch sphere, i.e. $\ket{\psi_1}=V(\vec{\kappa}_1)\ket{0}$, $\ket{\psi_2}=V(\vec{\kappa}_2)\ket{0}$,
$\ket{\psi}=V(\vec{\kappa}_3)\ket{0}$, with $V$ a general one-qubit unitary and each 
$\vec{\kappa}_i$ a 3-dimensional vector, resulting in a real $9$-dimensional $\vec{\zeta}_i$. 
This yields the two-qubit correlated measurement 
$\mathcal{M}(\vec{\zeta}_i)\otimes\mathcal{M}(\vec{\zeta}_i)$.

With these parts we construct the parameterization
\begin{equation}
	\mathcal{G}_i=\eta_0\mathcal{U}(\vec{\phi}_i) +  
	              \eta_1\mathcal{M}(\vec{\zeta}_i)\otimes \mathcal{M}(\vec{\zeta}_i),
\label{ansatz}
\end{equation}
with coefficients $\eta_i\in\mathbb{R}$ that satisfy $\eta_0+\eta_1=1$ because $\mathcal{G}_i$ is 
trace preserving. Note that here the tensor product symbol corresponds to combining two one-qubit channels to make 
a two-qubit channel, whereas in most of the paper it is used to link the column and row indices 
of a density matrix.
We construct the denoiser from the noisy channels $\tilde{\mathcal{G}}_i=\mathcal{N}\mathcal{G}_i$.
With this parameterization one denoiser channel has $17$ independent real parameters, 
such that a denoiser of depth $M$, i.e. consisting of $M$ brickwall layers, has $34M$ real parameters 
(we use one unique channel per half brickwall layer). For reference, a general channel has 
$544M$ parameters. 

\begin{figure}
        \centering
        \includegraphics[width=\columnwidth]{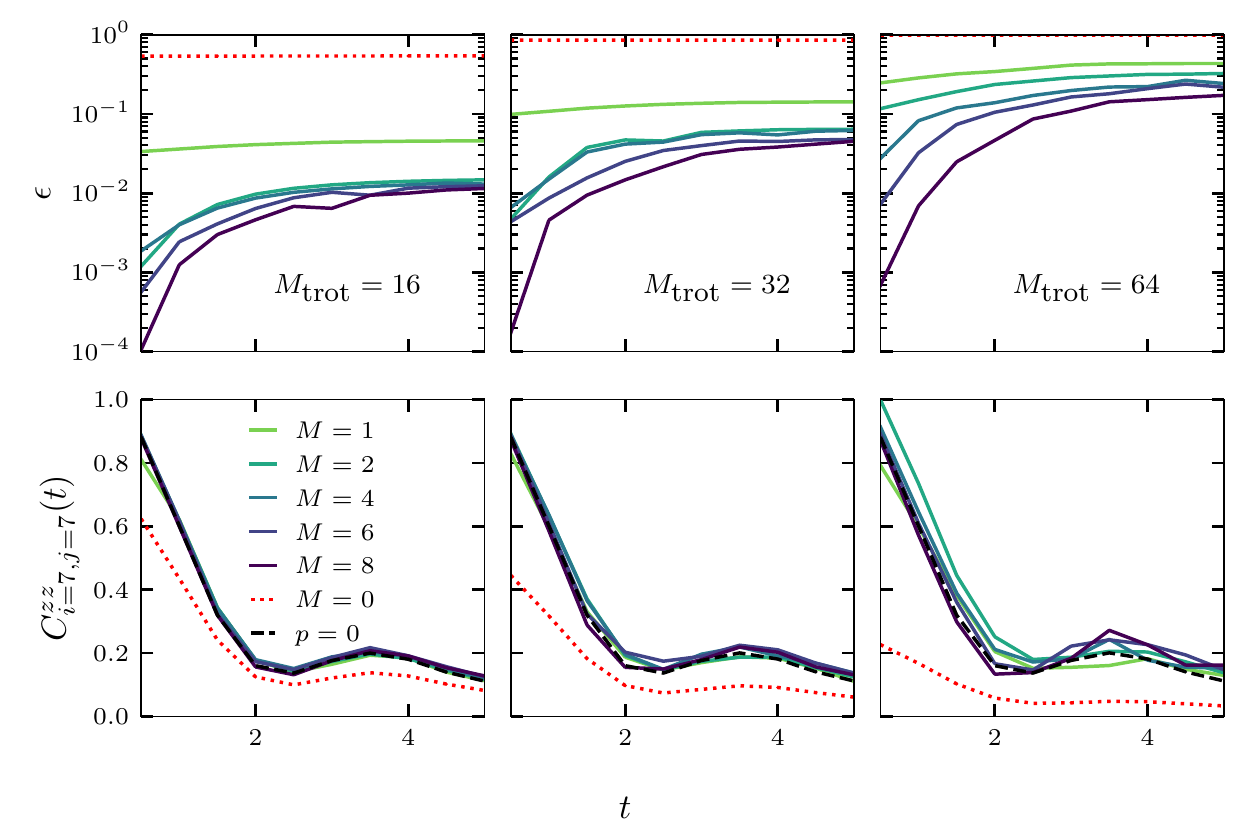}
        \caption{The normalized distance $\epsilon$ between the denoised Trotter supercircuit
                 $\tilde{\mathcal{D}}\tilde{\mathcal{C}}$ and the noiseless Trotter supercircuit 
                 $\mathcal{C}$ (top panels), at evolution times $t=0.5,1,...,5$, and the two-point 
                 $z$-spin correlator $C^{zz}_{i=L/2,j=L/2}(t)$ of a spin on the middle site at times 
                 $0$ and $t$ (bottom panels), for the infinite temperature initial state. We consider 
                 denoisers with depths $M=1,2,4,6,8$ and second-order Trotter circuits with depths 
                 $M_{\text{trot}}=16,32,64$. In the top panels we use a Heisenberg chain with $L=8$, 
                 and in the bottom panels with $L=14$, both with periodic boundary conditions.
                 All gates are affected by two-qubit depolarizing noise with $p=0.01$. 
                 The non-denoised results are labelled with $M=0$, and the noiseless values with $p=0$.}
        \label{rel_distance}
\end{figure}

To determine the mitigated expectation values we use the full expression
\begin{equation}
    \langle\hat{O}\rangle_{p=0} =\langle\langle \mathbb{1} | (\hat{O}\otimes \mathbb{1}) \mathcal{C} |\rho_0\rangle\rangle    
    \approx\langle\langle \mathbb{1}|(\hat{O}\otimes \mathbb{1})\tilde{\mathcal{D}}\tilde{\mathcal{C}}|\rho_0\rangle\rangle,
\label{O_super}
\end{equation}
where $|\rho_0\rangle\rangle$ is the initial state
and $|\mathbb{1}\rangle\rangle$ is the vectorized identity operator on the full Hilbert space.
To evaluate this on a quantum processor, we use the stochastic interpretation 
of (\ref{ansatz}) to resample (\ref{O_super}). In particular, from each channel (\ref{ansatz}) we 
get a unitary with probability $p_0=|\eta_0|/\gamma$ and a measurement followed by conditional preparation with probability $p_1=|\eta_1|/\gamma$. 
Here $\gamma=|\eta_0|+|\eta_1|$ is the sampling overhead, which characterizes the magnitude of the sign 
problem from negative $\eta_i$ \cite{endo_practical_2018, temme_error_2017, takagi_optimal_2021, piveteau_quasiprobability_2022, guo_noise_2022, jiang_physical_2021}. 
For quasiprobability distributions, i.e. with $\gamma>1$, 
every denoiser sample has an extra sign $\text{sgn}(\eta)=\prod_{g=1}^{N_{\mathcal{G}}}\text{sgn}(\eta_g)$, where 
$\text{sgn}(\eta_g)$ is the sign of the sampled coefficient of the $g$th channel. $\gamma=1$ means that all signs are positive.
Observables $\langle \hat{O}\rangle_{p=0}$ for the noiseless circuit are then approximated by resampling the 
observables from the denoiser ensemble \cite{temme_error_2017}
\begin{equation}
	\langle \hat{O}\rangle_{p=0}\approx\gamma\langle\text{sgn}(\eta)\hat{O}\rangle_{p}, 
\end{equation}
where $\gamma=\prod_{g=1}^{N_{\mathcal{G}}}\gamma_g$ is the overall sampling overhead, 
with $\gamma_g$ the overhead of the $g$th gate. 
Clearly, a large $\gamma$ implies a large variance of $\langle\hat{O}\rangle_{p=0}$ for a given number  
of samples, with accurate estimation requiring the cancellation of large signed terms.

The number of samples required to resolve this cancellation of signs is bounded by 
Hoeffding's inequality, which states that a sufficient number of samples to estimate 
$\langle\hat{O}\rangle_{p=0}$ with error $\delta$ at probability $1-\omega$ is bounded by 
$(2\gamma^2/\delta^2)\ln(2/\omega)$ \cite{takagi_optimal_2021}. Since $\gamma$ 
scales exponentially in $\gamma_g$, it is clear that a denoiser with large $M$ 
and $\gamma\gg1$ will require many samples. 

We observed that decompositions with $\gamma>1$ are crucial for an accurate denoiser. Restricting to $\gamma=1$ 
leads to large infidelity and no improvement upon increasing the number of terms in (\ref{ansatz}) 
or the depth $M$ of the denoiser. 
Simply put, probabilistic error cancellation of gate noise introduces 
a sign problem and it is crucial to find optimal parameterizations (\ref{ansatz}) which minimize $\gamma$ to make 
the approach scalable.
This issue arises in all high performance error mitigation schemes 
\cite{temme_error_2017, takagi_optimal_2021, jiang_physical_2021, berg_probabilistic_2022}, 
because the inverse of a physical noise channel is unphysical and cannot be represented as a positive sum over CPTP
maps. This is clearly visible in the spectra of the denoiser, which lies outside the unit circle (cf. Fig.
\ref{spectra}).
This makes the tunability of the number of gates in each denoiser sample a crucial ingredient, 
which allows control over the sign problem, because we can freely choose the $\eta_i$ in (\ref{ansatz}). 

For the parametrization \eqref{ansatz} of denoiser channels, we try to find a set 
of parameters for error mitigation by minimizing the normalized Frobenius distance 
between the noiseless and denoised supercircuits \cite{tepaske_optimal_2022}
\begin{equation}
    \epsilon=||\mathcal{C}-\tilde{\mathcal{D}}\tilde{\mathcal{C}}||^2_F/4^L,
    \label{eq:epsilon}
\end{equation}
which bounds the distance of output density matrices and becomes zero for perfect denoising.

We carry out the minimization of $\epsilon$ on a classical processor, 
using gradient descent with the differential programming algorithm from \cite{tepaske_optimal_2022}.  
Instead of explicitly calculating the accumulated global noise channel and subsequently inverting it,
we approximate the noiseless supercircuit $\mathcal{C}$ with the denoised supercircuit 
$\tilde{\mathcal{D}}\tilde{\mathcal{C}}$, effectively yielding a circuit representation $\mathcal{D}$ 
of the inverse noise channel.

\textit{Results.} --- 
To benchmark the denoiser we apply it to the second-order Trotter circuits of the spin-$1/2$ 
Heisenberg chain with periodic boundary conditions (PBC)
\begin{equation}
    H=\sum_{i=1}^L \left( \sigma^i_1\sigma^{i+1}_1+\sigma^i_2\sigma^{i+1}_2+\sigma^i_3\sigma^{i+1}_3 \right),
\end{equation}
where $\sigma^i_\alpha=(\mathbb{1}^i, \sigma^i_x, \sigma^i_y, \sigma^i_z)$ is the Pauli algebra 
acting on the local Hilbert space of site $i$. A second-order Trotter circuit for evolution time $t$ 
with depth $M_{\text{trot}}$ consists of $M_{\text{trot}}-1$ half brickwall layers with time step 
$t/M_{\text{trot}}$ and two layers with half time step \cite{tepaske_optimal_2022, paeckel_time_2019}. 
We consider circuits that are affected by uniform depolarizing noise with probability $p$ for simplicity, but our
approach can be used for any non-Clifford noise.
The two-qubit noise channel is
\begin{equation}
	\mathcal{N}=\left(1-\frac{16p}{15}\right)\mathbb{1} +  
	            \frac{p}{15}\bigotimes_{j=i}^{i+1}\left(\sum_{\alpha=0}^3\sigma^j_{\alpha}\otimes\sigma^{j*}_{\alpha}\right),
\end{equation}
which acts on neighboring qubits $i$ and $i+1$ and is applied to each Trotter and denoiser gate, and $p=0.01$ unless
stated otherwise.
We study circuits with depths $M_{\text{trot}}=16,32,64$ for evolution times 
$t=0.5,1,...,5$, and denoisers $\tilde{\mathcal{D}}$ with depths $M=1,2,4,6,8$.

In the top panels of Fig. \ref{rel_distance} we show $\epsilon$ \eqref{eq:epsilon} for 
a chain of size $L=8$ as a function of time $t$. 
Here it can be seen that even for $M_{\text{trot}}=32$ a denoiser with $M=1$ already improves $\epsilon$ 
by roughly an order of magnitude at all considered $t$. Depending on $M_{\text{trot}}$ and $t$, 
further increasing $M$ lowers $\epsilon$, with the biggest improvements occurring for high precision Trotter circuits
with large depth $M_{\text{trot}}=64$ and short time $t=0.5$, where the Trotter gates are closer to the identity
than in the other cases. 
At the other extreme, for $M_{\text{trot}}=16$ the improvements are relatively small upon increasing $M>2$. 
In all cases the denoiser works better at early times than at late times, again indicating that it is easier 
to denoise Trotter gates that are relatively close to the identity. 

\begin{figure}
        \centering
        \includegraphics[width=\columnwidth]{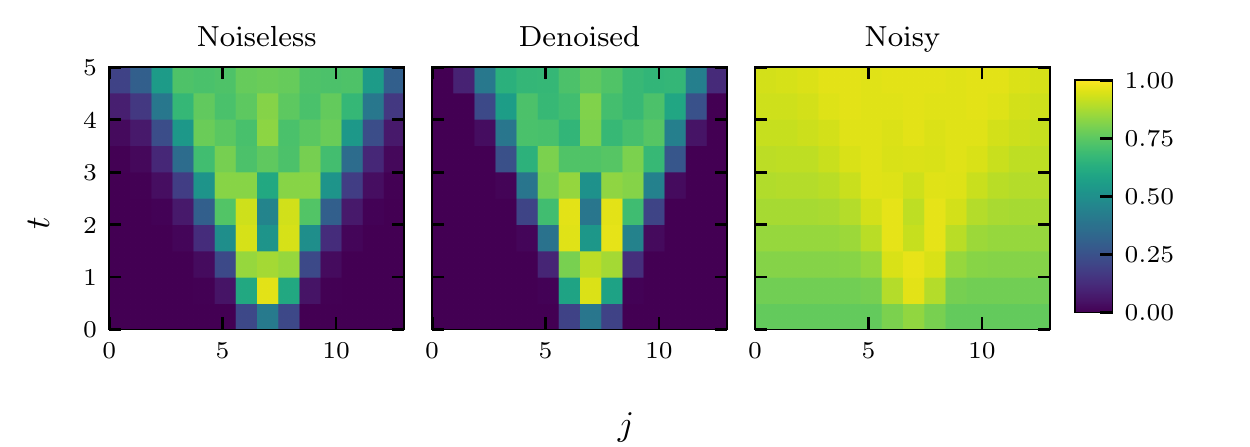}
        \caption{The out-of-time-ordered correlator $C^{\text{otoc}}_{i=L/2,j}(t)$ as a function 
                 of the operator position $j$ and time $t$, for the infinite temperature initial state,
                 for a denoised second-order Trotter supercircuit with Trotter depth $M_{\text{trot}}=32$ 
                 and denoiser depth $M=2$. We consider evolution times $t=0.5,1,...,5$, for the periodic 
                 $L=14$ Heisenberg chain that is affected by two-qubit depolarizing noise with $p=0.01$. 
                 }
        \label{otoc}
\end{figure}

To probe the accuracy of the denoiser on quantities that do not enter the optimization, as 
a first test we consider the two-point correlator between spins at different times \cite{dupont_universal_2020}
\begin{equation}
	C^{zz}_{ij}(t)=\langle\langle \mathbb{1}|(\sigma^z_{i}\otimes \mathbb{1})\tilde{\mathcal{D}}\tilde{\mathcal{C}}(t)(\sigma^z_{j}\otimes \mathbb{1}) |\mathbb{1}\rangle\rangle/2^L,
	\label{eq:corr}
\end{equation}
where we have chosen the infinite temperature initial state, 
and $\tilde{\mathcal{C}}(t)$ is the Trotter supercircuit for time $t$.
In the bottom panels of Fig. \ref{rel_distance} we show $C^{zz}_{i=L/2,j=L/2}(t)$ for the supercircuits from the upper panels, now for a $L=14$ chain. 
Here we see that at $M_{\text{trot}}=16$ we can retrieve the noiseless values already with $M=1$, but 
that increasing $M_{\text{trot}}$ makes this more difficult. At $M_{\text{trot}}=64$ we see larger deviations, and 
improvement upon increasing $M$ is less stable, but nonetheless we are able to mitigate errors to a large extent.

As a further test, we compute the out-of-time-ordered correlator (OTOC) \cite{zhang_information_2019, tepaske_optimal_2022, 
hemery_matrix_2019, luitz_information_2017, maldacena_bound_2016, larkin_quasiclassical_1969}
\begin{multline}
	C^{\text{otoc}}_{ij}(t) =	\textrm{Re}\langle\langle \mathbb{1}|(\sigma^{z\dagger}_j\otimes \mathbb{1})\tilde{\mathcal{D}}\tilde{\mathcal{C}}(-t) \\ 
	(\sigma_i^z\otimes\sigma_i^{z*})\tilde{\mathcal{D}}\tilde{\mathcal{C}}(t) (\sigma^z_j\otimes \mathbb{1})|\mathbb{1}\rangle\rangle/2^L.
	\label{eq:otoc}
\end{multline}
In Fig. \ref{otoc} we show the results for $i=L/2$, for a Trotter 
circuit with depth $M_{\text{trot}}=32$ and a denoiser with depth $M=2$. Here we see that a denoiser with 
$M\ll M_{\text{trot}}$ is able to recover the light-cone of correlations, which are otherwise buried by the noise.
In the Supplementary Material we consider how the denoiser performs at different noise levels $p$, and how 
the denoised supercircuits perform under stacking. There we also calculate domain wall magnetization 
dynamics, and show the distribution of the optimized denoiser parameters and the 
sampling overhead associated to the denoiser as a whole. 

\begin{figure}
        \centering
        \includegraphics[width=\columnwidth]{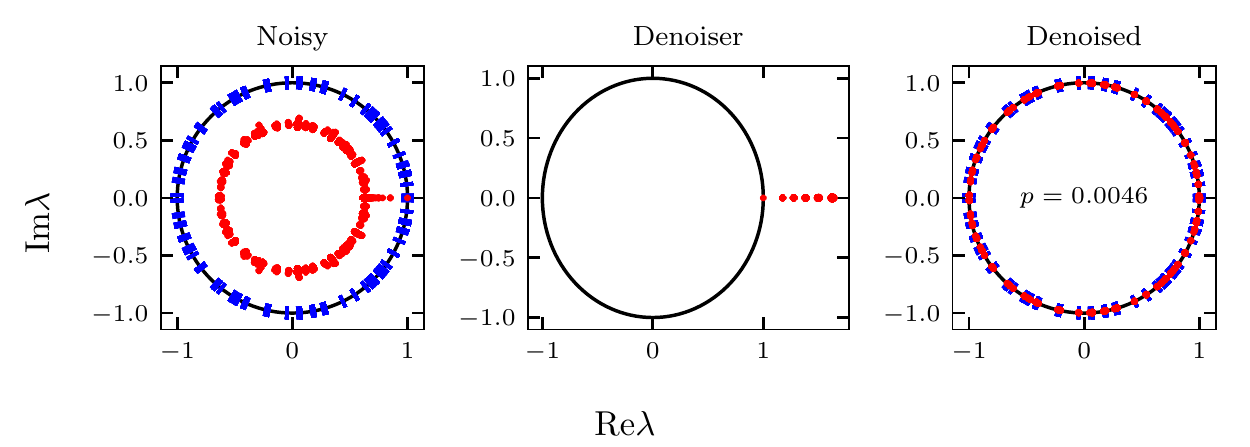}
        \caption{The complex eigenvalues $\lambda$ of the noisy second-order Trotter supercircuit with $M_{\text{trot}}=16$ at time $t=1$ (left), 
                 the corresponding optimized denoiser with $M=4$ (center), and the denoised Trotter 
                 supercircuit (right). The Trotter circuit is for a $L=6$ Heisenberg model with PBC,
                 and all two-qubit channels are affected by depolarizing noise with $p=0.0046$.
                 The unit circle, on which unitary eigenvalues must lie, is shown in black, and the noiseless 
                 eigenvalues are shown as blue bars. It is evident that the denoiser recovers all the noiseless 
                 eigenvalues from the noisy circuit.}
        \label{spectra}
\end{figure}

In Fig. \ref{spectra} we show the eigenvalues of 
the noisy supercircuits for a noisy second-order Trotter supercircuit with $M_{\text{trot}}=16$ at $t=1$ (left), 
the corresponding optimized denoiser with $M=4$ (center), and the denoised supercircuit (right). 
The eigenvalues $\lambda$ of a unitary supercircuit lie on the unit circle, and in the presence 
of dissipation they are pushed to the center. We see that the spectrum of the denoiser lies outside the unit circle,
making it an unphysical channel which cures the effect of the noise on the circuit, such that the spectrum of the
denoised circuit is pushed back to the unit circle. The noiseless eigenvalues are shown as blue bars, 
making it clear that the denoiser is able to recover the noiseless eigenvalues from the noisy circuit.
In the Supplementary Material we 
show the spectra for a $p=0.036$ denoiser, where we observe a clustering of eigenvalues reminiscent of Refs.
\cite{wang_hierarchy_2020,li_random_2022,sommer_many-body_2021}. There we also investigate the 
channel entropy of the various supercircuits \cite{zhou_operator_2017, roga_entropic_2013}.

\textit{Conclusion.} --- 
We have introduced a probabilistic error cancellation scheme, where a classically determined denoiser 
mitigates the accumulated noise of a (generally non-Clifford) local noise channel. The required number of 
mitigation gates, i.e. the dimensionality of the corresponding quasiprobability distribution, is tunable 
and the parameterization of the corresponding channels provides control over the sign problem that 
is inherent to probabilistic error cancellation. 
We have shown that a denoiser with one layer can already significantly mitigate 
errors for second-order Trotter circuits with up to $64$ layers. 

This effectiveness of low-depth compressed circuits for denoising, in 
contrast with the noiseless time evolution operator compression from \cite{tepaske_optimal_2022}, 
can be understood from the non-unitarity of the denoiser channels. In particular, 
measurements can have non-local effects, since the measurement of a single qubit can reduce some highly entangled 
state (e.g. a GHZ state) to a product state, whereas in unitary circuits the spreading of correlations 
forms a light-cone.

To optimize a denoiser with convenience at $L>8$, the optimization can be formulated in terms of matrix 
product operators \cite{tepaske_optimal_2022, guo_quantum_2022} or channels \cite{filippov_matrix_2022}, which is convenient 
because the circuit calculations leading to the normalized distance $\epsilon$ and its gradient are easily formulated in terms of tensor 
contractions and singular value decompositions \cite{tepaske_optimal_2022, noh_efficient_2020}.
This provides one route to a practical denoiser, which is relevant because the targeted noiseless circuit and the 
accompanying noisy variant in (\ref{eq:epsilon}) need to be simulated classically, confining the optimization procedure 
to limited system sizes with an exact treatment or limited entanglement with tensor networks. 
Nonetheless, we can use e.g. matrix product operators to calculate (\ref{eq:epsilon}) for 
some relatively small $t$, such that the noiseless and denoised supercircuits in (\ref{eq:epsilon}) have 
relatively small entanglement, and then stack the final denoised supercircuit on a quantum processor to generate 
classically intractable states. Analogously, we can optimize the channels exactly at some classically tractable size 
and then execute them on a quantum processor with larger size. Both approaches are limited by the light-cone of many-body 
correlations, as visualized in Fig. \ref{otoc}, because finite-size effects appear when the light-cone width becomes comparable with system size.

\section*{Acknowledgments}

We are grateful for extensive discussions with Dominik Hahn.
This project was supported by the Deutsche Forschungsgemeinschaft (DFG) through the 
cluster of excellence ML4Q (EXC 2004, project-id 390534769). 
We also acknowledge support from the QuantERA II Programme that has received funding from 
the European Union’s Horizon 2020 research innovation programme (GA 101017733), and
from the Deutsche Forschungsgemeinschaft through the project DQUANT (project-id 499347025).

\clearpage

\section{Supplementary material}

\subsection{Denoiser performance at various noise levels}\label{app:p_dependency}

\begin{figure}
        \centering
        \includegraphics[width=\columnwidth]{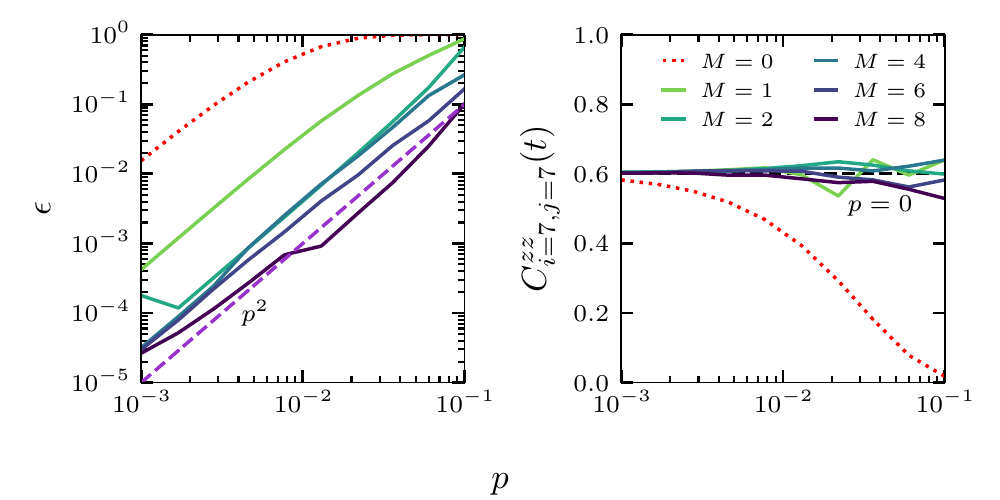}
        \caption{The normalized distance $\epsilon$ (left) and
                 $z$ spin correlator $C^{zz}_{i=L/2,j=L/2}$ (right), for a second-order Trotter supercircuit of depth
                 $M_{\text{trot}}=16$ for time $t=1$, affected by various two-qubit depolarizing errors $p$.
                 We compare the values obtained with and without a denoiser, i.e. $M>0$ and $M=0$,
                 to the noiseless values ($p=0$).
                 The denoiser is affected by the same noise as the Trotter circuit. We
                 consider denoisers with depths $M=1,2,4,6,8$, and we use a $L=8$ Heisenberg chain
                 with PBC for the normalized distance, while for the correlator we use $L=14$.}
        \label{p_dependency}
\end{figure}

To probe how the denoiser performs at different noise strengths $p$, we take a second-order
Trotter supercircuit of depth $M_{\text{trot}}=16$ for the time evolution of the wave function to
time $t=1$, and optimize the
denoiser at various noise strengths in the interval $p\in\left[10^{-3},10^{-1}\right]$.
In Fig. \ref{p_dependency} we show the normalized distance (left panel) and the $z$ spin correlator (right), for denoiser depths $M=1,2,4,6,8$.
For comparison, we show the results for the noisy limit, i.e. without a denoiser ($M=0$, red dashed),
and for the exact limit without noise ($p=0$, black dashed).

The error of the entire circuit $\epsilon$ improves with denoiser depth $M$ for the full range of $p$, and depends
roughly quadratically on $p$. This is illustrated with the purple dashed line in the left panel of Fig. \ref{p_dependency}.
It is interesting to observe that even for larger noise strength $p$, the local observable
$C^{zz}$ improves significantly even with denoisers of depth $M=1$. For large noise strengths, we generally see
that the optimization of the denoiser becomes difficult, leading to nonmonotonic behavior as a function of $p$,
presumably because we do not find the global optimum of the denoiser.

\subsection{Supercircuit spectra}

\begin{figure}
        \centering
        \includegraphics[width=\columnwidth]{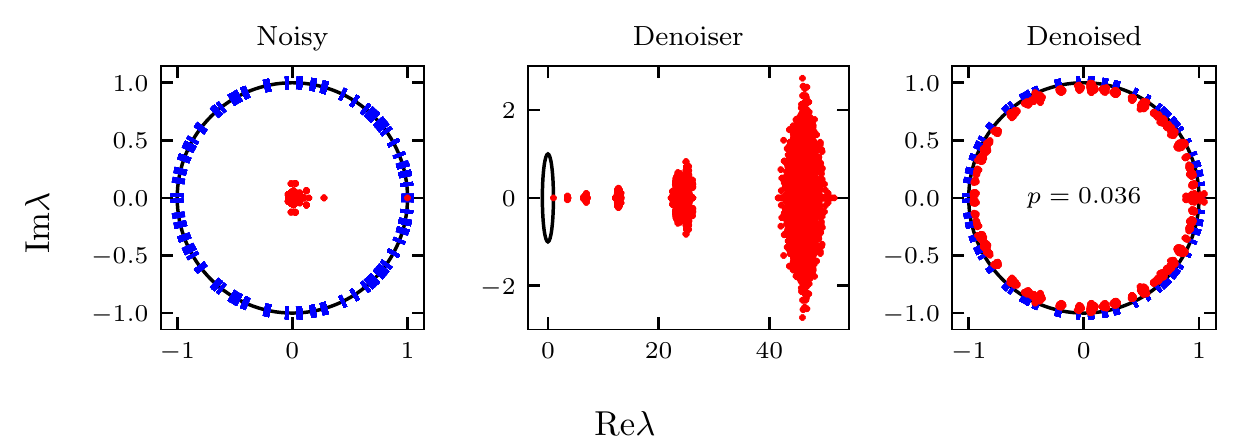}
        \caption{The complex eigenvalues $\lambda$ of the noisy second-order Trotter supercircuit with $M_{\text{trot}}=16$ at time $t=1$ (left),
                 the corresponding optimized denoiser with $M=4$ (center), and the denoised Trotter
                 supercircuit (right). The Trotter circuit is for a $L=6$ Heisenberg model with PBC,
                 and all two-qubit channels are affected by depolarizing noise with $p=0.036$.
                 The unit circle, on which unitary eigenvalues must lie, is shown in black, and the noiseless
                 eigenvalues are shown as blue bars. It is clear that the denoiser recovers with high accuracy the
                 noiseless eigenvalues from the noisy circuit.}
        \label{spectra_p0.036}
\end{figure}

It is interesting to analyze the spectra of the supercircuits considered in this work. As mentioned in the main text,
the spectrum of the ideal, unitary supercircuit $\mathcal{C}$ lies on the unit circle. The comparison to this case is
therefore instructive. In the main text, we showed an example of the spectra in Fig. 4 for moderate noise
strength. Here, we show additional data for stronger noise $p=0.036$
in Fig. \ref{spectra_p0.036} for a denoiser with $M=4$ layers, optimized to mitigate errors for a
second-order Trotter supercircuit with $M_{\text{trot}}=16$ layers at time $t=1$.

The eigenvalues $\lambda$ of the noisy supercircuit $\mathcal{\tilde C}$ are clustered close to zero, far away from the
unit circle (except for $\lambda=1$), showing that the circuit is strongly affected by the noise.
To mitigate the impact of the noise, the denoiser consequently has to renormalize the spectrum strongly. If it
accurately represents the inverse of the global noise channel, its spectrum has to lie far outside the unit circle, which
is the case. Interestingly, we observe a clustering of eigenvalues which is reminiscent to the spectra found
in \cite{wang_hierarchy_2020,sommer_many-body_2021,li_random_2022}. By comparison to these works, we suspect that this
is due to the local nature of the denoiser, and warrants further investigation.

The right panel of Fig. \ref{spectra_p0.036} shows the result of the denoiser, pushing the eigenvalues back to the unit
circle, nearly with the exact same distribution along the circle as the noiseless eigenvalues (blue bars).
Due to the strong noise, this is not achieved perfectly, and it is clear that this cannot work in principle if
the global noise channel has a zero eigenvalue.

\subsection{Supercircuit entropies}\label{app:p_dependency}

\begin{figure}
        \centering
        \includegraphics[width=\columnwidth]{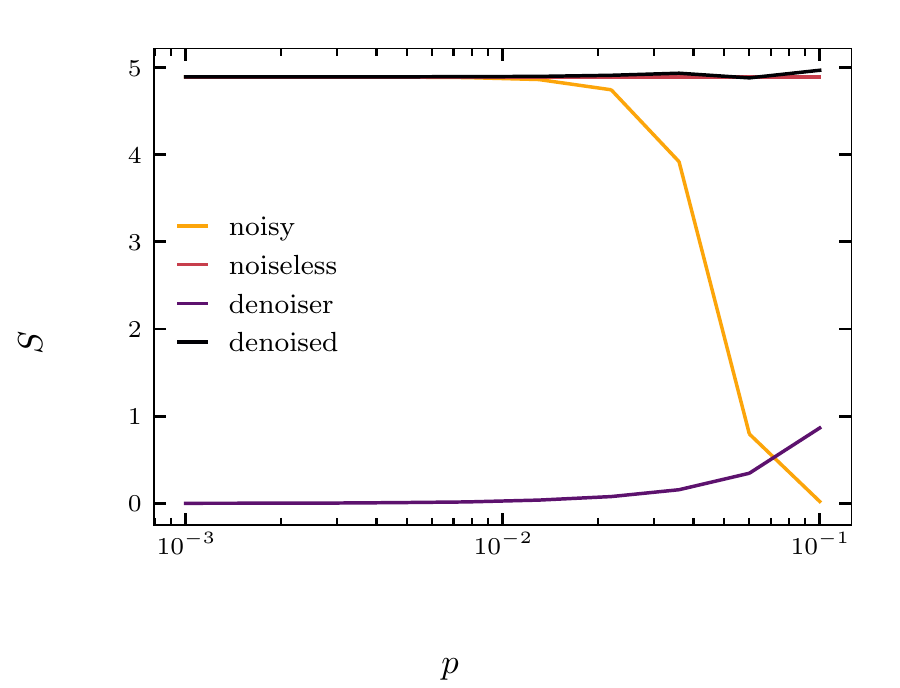}
        \caption{The half-chain channel entanglement entropy $S$ at different two-qubit depolarizing noise
                 strengths $p$, for a second-order Trotter supercircuit with $M_{\text{trot}}=16$ and
                 $t=2$, for a $M=4$ denoiser. The Trotter circuit is for a Heisenberg model with
                 PBC of size $L=6$. The different curves correspond to the different supercircuits,
                 i.e. the noisy supercircuit, the denoiser, the corresponding denoised supercircuit,
                 and the noiseless variant.}
        \label{ee_dependency}
\end{figure}

The complexity of an operator can be quantified by its operator entanglement entropy \cite{zhou_operator_2017}.
Here we calculate the half-chain channel entanglement entropy $S$ \cite{roga_entropic_2013} of the noiseless $\mathcal
C$, noisy $\tilde{\mathcal{C}}$, denoiser $\tilde{\mathcal{D}}$, and denoised $\tilde{\mathcal{D}} \tilde{\mathcal{C}}$ supercircuits. We define $S$ as the entanglement entropy of the
state that is related to a supercircuit $\mathcal{C}$ via the Choi-Jamio\l kowski
isomorphism, i.e. $\psi_{\mathcal{C}}=\chi_{\mathcal{C}}/N$, where
the process matrix $\chi^{ab,cd}_{\mathcal{C}}=\mathcal{C}^{ac,bd}$ is simply a reshaped
supercircuit and $N$ ensures normalization. Then we have $S=-\tr\left[\psi_{\mathcal{C}}\ln\psi_{\mathcal{C}}\right]$.
This entropy measure is a particular instance of the ``exchange entropy'', which characterizes
the information exchange between a quantum system and its environment \cite{roga_entropic_2013}.

In Fig. \ref{ee_dependency} we plot the various $S$ for a second-order Trotter circuit with $M_{\text{trot}}=16$ at
$t=2$, for a denoiser with $M=4$, both affected by two-qubit depolarizing noise with $p\in[10^{-3},10^{-1}]$. The Trotter
circuit is for a Heisenberg model with $L=6$ and PBC. We see that at large $p$, the noise destroys entanglement in the noisy
supercircuit, and that the denoiser $S$ increases to correct for this, such that the denoised
supercircuit recovers the noiseless $S$.

\subsection{Stacking denoised supercircuits}\label{app:stacking}

\begin{figure}
        \centering
        \includegraphics[width=\columnwidth]{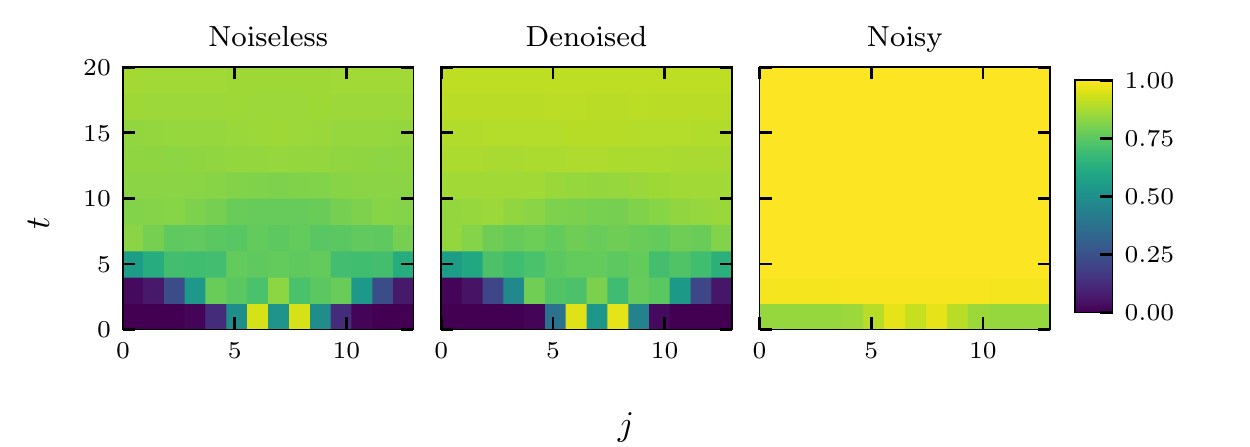}
        \caption{The out-of-time-ordered correlator $C^{\text{otoc}}_{i=L/2,j}(t)$ as a function
                 of the operator position $j$ and stacked time $t$, for the infinite temperature initial state,
                 for a denoised second-order Trotter supercircuit with Trotter depth $M_{\text{trot}}=32$
                 and denoiser depth $M=2$. It is optimized at $t=2$ and stacked up to ten times.
                 The calculations are for the periodic $L=14$ Heisenberg chain that is affected by
                 two-qubit depolarization with $p=0.01$. The denoiser is affected by the same noise.}
        \label{stacking_otoc}
\end{figure}

\begin{figure}
        \centering
        \includegraphics[width=\columnwidth]{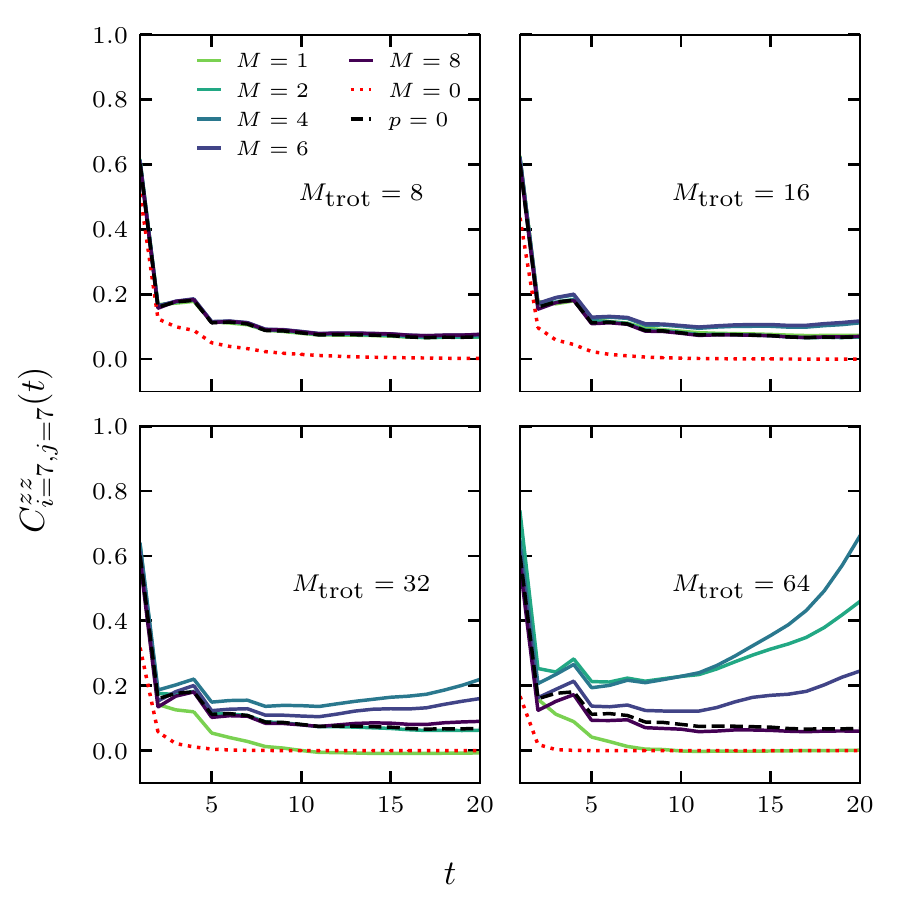}
        \caption{The two-point $z$-spin correlator $C^{zz}_{i=L/2,j=L/2}(t)$ of a spin on
                 the middle site at times $0$ and $t$, for the infinite temperature initial state,
                 for denoised second-order Trotter supercircuits that are optimized at evolution time
                 $t=1$ and then stacked up to twenty times.
                 We use Trotter depths $M_{\text{trot}}=8,16,32,64$ and denoiser depths $M=1,2,4,6,8$.
                 The calculations were performed for a periodic Heisenberg model with $L=14$ and PBC,
                 affected by two-qubit depolarizing noise with strength $p=0.01$, which also affects the denoiser.
                 The non-denoised results are labelled with $M=0$, and the noiseless results with $p=0$.
                 The panels are arranged as $M_{\text{trot}}=8,16,32,64$ for top left, top right, bottom left, bottom right, respectively.}
        \label{stacking_corr}
\end{figure}

Here we investigate how denoised supercircuits perform upon repeated application.
We optimize the denoiser for a Trotter supercircuit for a fixed evolution time $t$. Then, to reach
later times, we stack the denoised supercircuit $n$ times to approximate the evolution up to time $nt$:
\begin{equation}
    \mathcal{C}(nt) \approx \left( \tilde{\mathcal{D}}(t) \tilde{\mathcal{C}}(t) \right)^{n}
    \label{eq:stacking}
\end{equation}
In Fig. \ref{stacking_corr} we stack a denoised $t=1$ supercircuit up to $n=20$ times and calculate the correlation function,
defined in the main text, for the middle site.
We consider Trotter depths $M_{\text{trot}}=8,16,32,64$ and denoiser depths $M=1,2,4,6,8$, for a $L=14$ Heisenberg
chain with $p=0.01$ depolarizing two-qubit noise. The noisy results correspond to $M=0$ and the noiseless results to $p=0$.
In Fig. \ref{stacking_otoc} we calculate the OTOC, defined in the main text, with stacked time evolution for a denoised $t=2$ supercircuit with
$M_{\text{trot}}=32$ and $M=2$, stacked up to ten times.
We see that the stacked supercircuit performs very well, and the additional precision obtained by using deep denoisers
($M=8$) pays off for long evolution times, where we see convergence to the exact result (black dashed lines in
Fig. \ref{stacking_corr}) as a function of $M$.

\subsection{Distribution of optimized ZZ channels}\label{app:alpha_dist}

\begin{figure}
        \centering
        \includegraphics[width=\columnwidth]{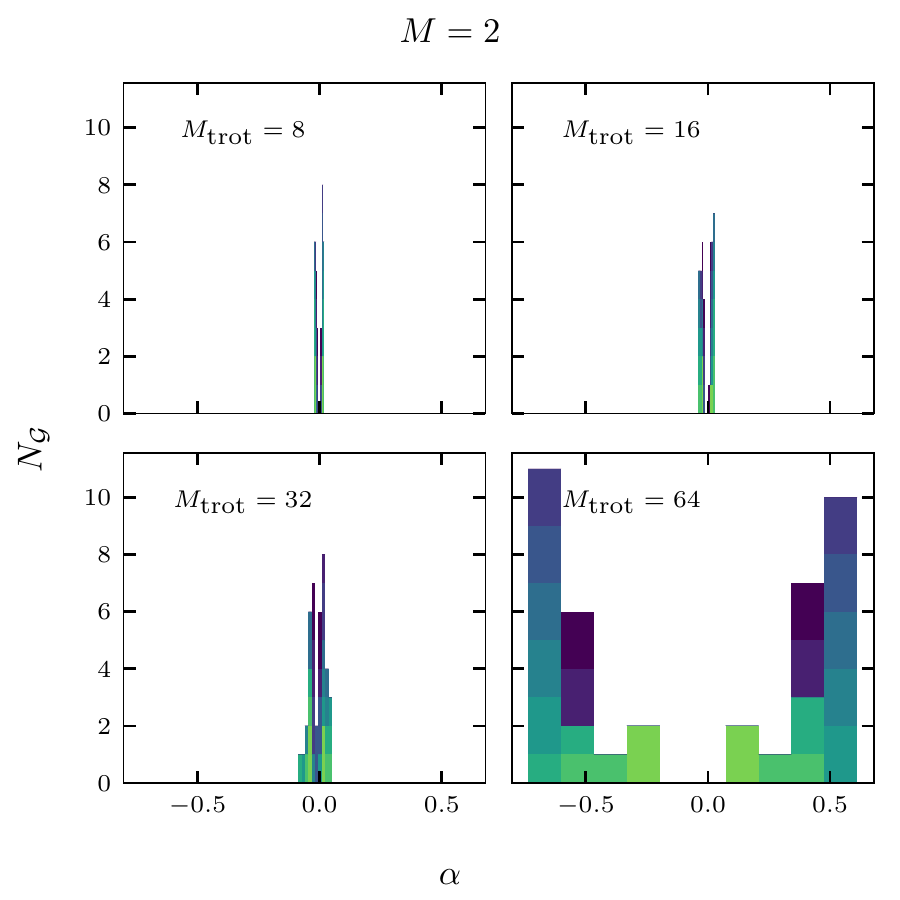}
        \includegraphics[width=\columnwidth]{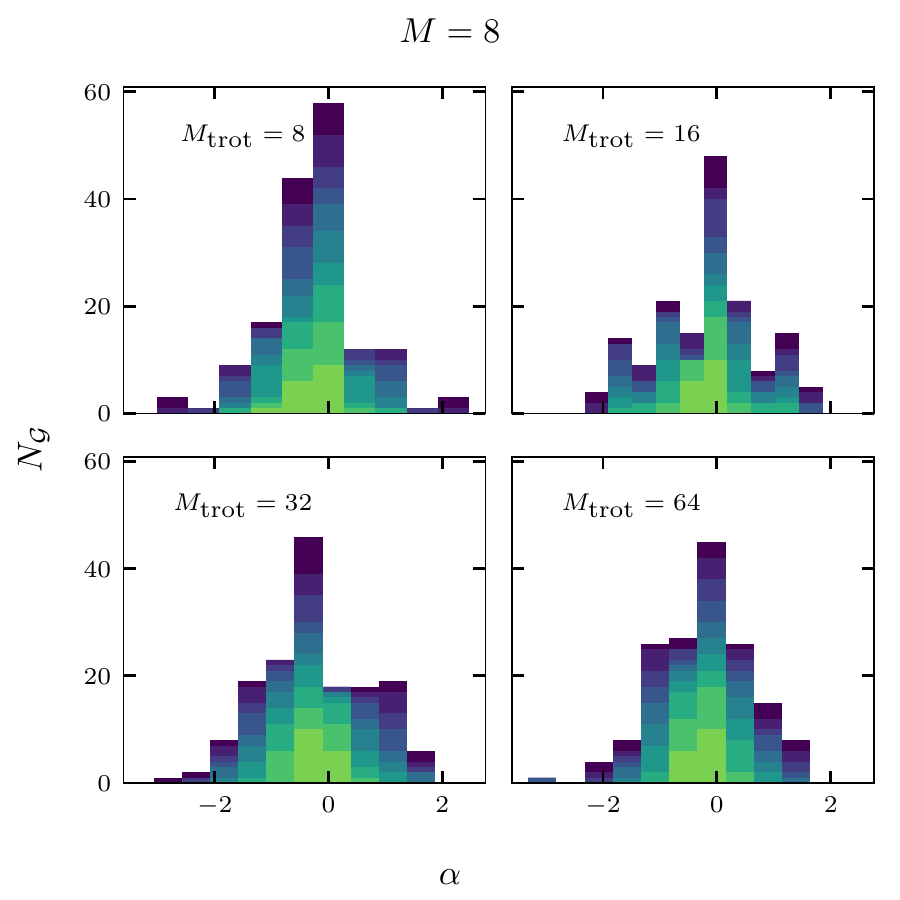}
        \caption{The distribution of the ZZ angle $\alpha$ of $M=2$ denoisers
                 (top panels) and $M=8$ denoisers (bottom panels), with the lightest color corresponding
                 to the denoiser for the Trotter supercircuit with $t=0.5$, and the darkest color with $t=5$.
                 As usual, we consider the Heisenberg model on a periodic chain, and second-order
                 Trotter supercircuits with depths $M_{\text{trot}}=8,16,32,64$, which together with the denoiser
                 is affected by a two-qubit depolarizing noise with $p=0.01$.
                 The panels are arranged as $M_{\text{trot}}=8,16,32,64$ for top left, top right, bottom left, bottom right, respectively.}
        \label{alphas}
\end{figure}

The costliest and most noise-susceptible operation is the two-qubit $ZZ$ rotation with angle $\alpha$, which
is the foundation of the unitary piece in our channel parameterization, defined in the main text.
For completeness, we here present the $\alpha$ angles of the optimized denoisers.
The results are shown in Fig. \ref{alphas}, which contains histograms for the channel count $N_{\mathcal{G}}$
versus $\alpha$. The histograms are stacked, with the lightest color
corresponding to the angles of the denoiser at $t=0.5$ and the darkest at $t=5$. The top four panels are for
a denoiser with $M=2$ and the bottom four with $M=8$. We consider $M_{\text{trot}}=8,16,32,64$.
We see that in both cases the distribution widens upon increasing $M_{\text{trot}}$, indicating that the
unitary channels start deviating more from the identity. Moreover, while the $M=2$ denoisers in all
cases except $M_{\text{trot}}=64$ have ZZ contributions close to the identity, this is clearly not the case for $M=8$.

\subsection{Sampling overhead of optimized denoisers}\label{app:nu}

\begin{figure}
        \centering
        \includegraphics[width=\columnwidth]{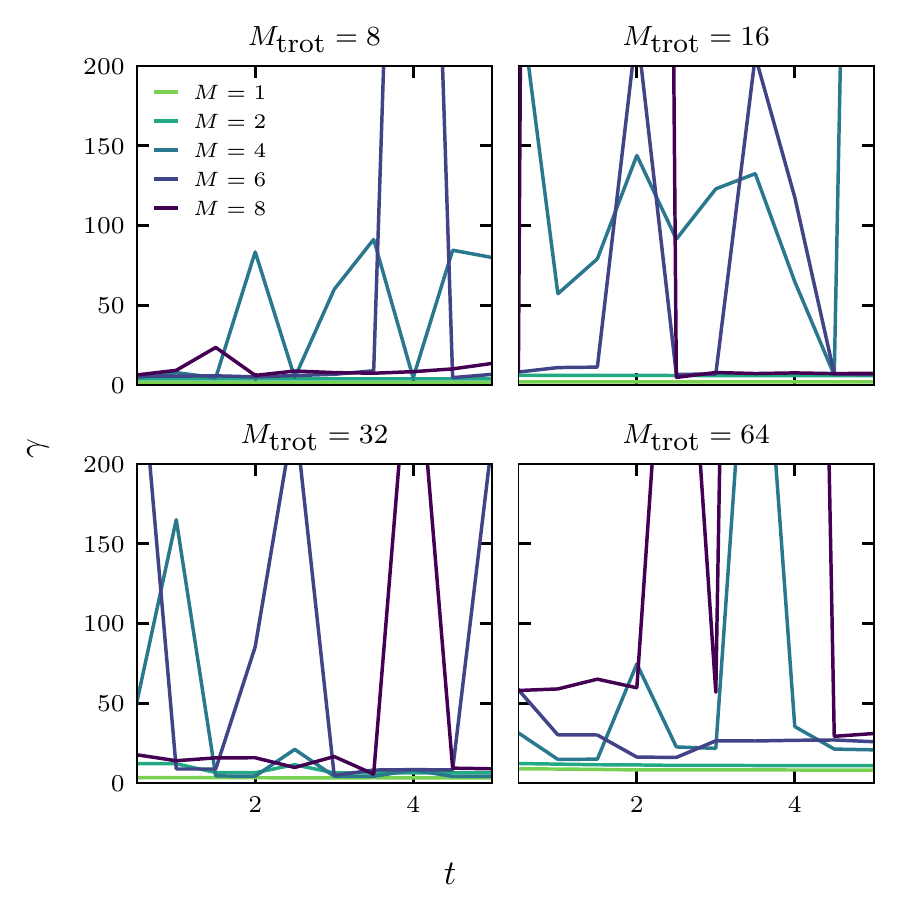}
        \caption{The sampling overhead $\gamma$ of the optimized denoisers from
                 Fig. 2 of the main text, with denoiser depths $M=1,2,4,6,8$ and Trotter
                 depths $M_{\text{trot}}=8,16,32,64$ at times $t=0.5,1,...,5$, for the Heisenberg model
                 on a chain with PBC affected by two-qubit depolarizing noise with $p=0.01$.
                 The panels are arranged as $M_{\text{trot}}=8,16,32,64$ for top left, top right, bottom left, bottom right, respectively.}
        \label{sampling_overhead}
\end{figure}

For simplicity, we did not focus on obtaining denoisers with the smallest sampling overhead $\gamma$, which
is required to minimize the sign problem and hence ease the sampling of mitigated quantities.
Instead, we let the optimization freely choose the $\eta_i$ in the denoiser parameterization, as defined in
the main text. In
Fig. \ref{sampling_overhead} we show the sampling overhead of the denoisers from
Fig. 2 of the main text. We see that for $M=1$ and $M=2$ the sampling overhead is relatively small and
uniform across the different $t$, whereas for $M>2$ the optimization sometimes yields a denoiser with large $\gamma$ and
other times with small $\gamma$. This could be related to the difference in $\alpha$ distributions from Fig. \ref{alphas}.
The large fluctuations of $\gamma$ appears to stem from the difficulty in finding optimal deep denoisers, and our
optimization procedure likely only finds a local minimum in these cases.

\subsection{Domain wall magnetization}\label{app:spectra}

\begin{figure}
        \centering
        \includegraphics[width=\columnwidth]{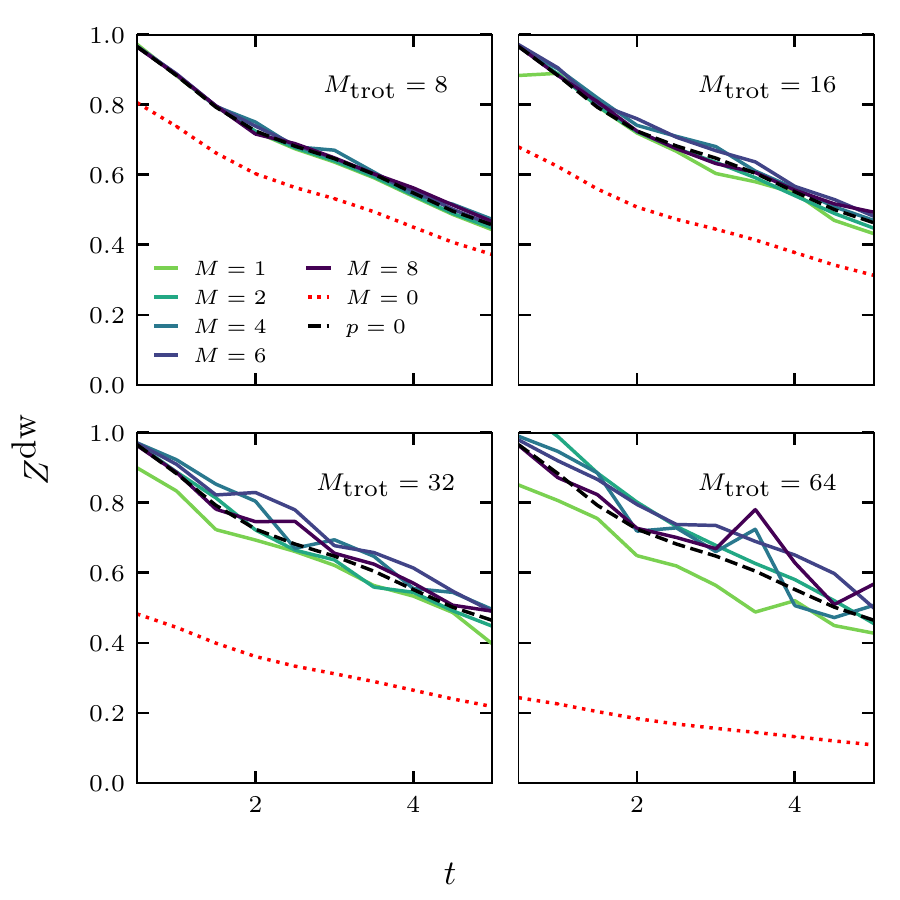}
        \caption{The domain wall magnetization $Z^{\text{dw}}$ after evolving a periodic density
                 wall $\ket{\text{dw}}\ket{\text{dw}^*}$ with the denoised second-order Trotter supercircuits $\tilde{\mathcal{D}}\tilde{\mathcal{C}}$
                 from Fig. 2 of the main text. These supercircuits have
                 various Trotter depths $M_{\text{trot}}=8,16,32,64$, denoiser depths $M=1,2,4,6,8$, and
                 evolution times $t=0.5,1,...,5$, for the periodic $L=14$ Heisenberg chain that is affected by
                 two-qubit depolarizing noise of strength $p=0.01$. The denoiser is affected by the same noise.
                 The non-denoised results are labelled with $M=0$ and the noiseless results with $p=0$.
                 The panels are arranged as $M_{\text{trot}}=8,16,32,64$ for top left, top right, bottom left, bottom right, respectively.
                 We see that the denoiser allows us to recover the noiseless behavior.}
        \label{dw_mag}
\end{figure}

As another test of the denoiser performance we evolve the periodic $z$-spin domain wall
$\ket{\text{dw}}=\bigotimes_{i=1}^{L/2}\ket{1}\otimes\bigotimes_{i=L/2}^{L}\ket{0}$ and consider
the domain wall magnetization
\begin{equation}
        Z^{\text{dw}}(t)=\sum_{i=1}^{L}(-1)^{\lfloor 2i/L \rfloor}\langle\langle \mathbb{1}|(\sigma^z_i\otimes \mathbb{1})\tilde{\mathcal{D}}\tilde{\mathcal{C}}(t)\ket{\text{dw}}\ket{\text{dw}^*}.
        \label{eq:mag}
\end{equation}
Here $\tilde{\mathcal{C}}(t)$ is the Trotter supercircuit for time $t$. In Fig. \ref{dw_mag} we show $Z^{\text{dw}}$
for the circuits from Fig. 2 of the main text, calculated for a $L=14$ chain.
As in our other tests, we see that at $M_{\text{trot}}=8$ we can recover the noiseless values already with $M=1$, and
that increasing $M_{\text{trot}}$ makes this more difficult. At $M_{\text{trot}}=64$ we see larger deviations, and
improvement upon increasing $M$ is less stable, but nevertheless we are able to mitigate errors to a large extent.

\clearpage

\bibliography{references}

\end{document}